\documentclass[prl,showpacs,twocolumn,amsmath,amssymb,groupadress,superscriptaddress]{revtex4}
\usepackage{graphicx}% Include figure files
\begin{document}

\newcommand{\REBa}{RBaCo$_4$O$_7$}
\newcommand{\YBa}{YBaCo$_4$O$_7$}
\newcommand{\YbBa}{YbBaCo$_4$O$_7$}
\newcommand{\Kag}{Kagom\'{e}}
\newcommand{\Deg}{$^{\circ}$}
\title{Competing magnetic interactions in the extended \Kag\ system \YBa}

\author{L.C. Chapon}
\affiliation{ISIS facility, Rutherford Appleton Laboratory-CCLRC,
Chilton, Didcot, Oxfordshire, OX11 0QX, United Kingdom. }
\author{P.G. Radaelli}
\affiliation{ISIS facility, Rutherford Appleton Laboratory-CCLRC,
Chilton, Didcot, Oxfordshire, OX11 0QX, United Kingdom. }
\affiliation{Dept. of Physics and Astronomy, University College
London, Gower Street, London WC1E 6BT, United Kingdom}
\author{H. Zheng}
\affiliation{Materials Science Division, Argonne National
Laboratory, Argonne, IL  60439}
\author{J.F. Mitchell}
\affiliation{Materials Science Division, Argonne National
Laboratory, Argonne, IL  60439}
\date{\today}% It is always \today, today,
             %  but any date may be explicitly specified

\begin{abstract}
\YBa\ belongs to a new class of geometrically frustrated magnets
like the pyrochlores, in which Co-spins occupy corners of
tetrahedra. The structure can be viewed as an alternating stacking
of \Kag\ and triangular layers. Exactly half of the triangular units
of the \Kag\ plane are capped by Co ions to form columns running
perpendicular to the \Kag\ sheets. Neutron powder diffraction
reveals a broad temperature range of diffuse magnetic scattering,
followed by long range magnetic ordering below 110K. A unique
low-temperature magnetic structure simultaneously satisfies an S=0
arrangement in the uncapped triangular units and antiferromagnetic
coupling along the columns. A spin reorientation above 30K tracks
the relative strengths of the in-plane and out-of-plane
interactions.
\end{abstract}

\pacs{25.40.Dn, 75.25.+z, 77.80.-e}% PACS, the Physics and Astronomy
                             % Classification Scheme.

\maketitle Magnetic frustration has attracted considerable interest
over the last 50 years\cite{Ramirez,Schiffer,Greedan}. Whether the
frustration arises from competing exchange interactions or the
peculiar geometry of the spin lattice, it invariably leads to
unconventional magnetic states at low temperatures. Frustrated
systems commonly exhibit the persistence of strong spin fluctuations
at low temperatures and, at least, partial suppression of the
magnetic order. Various spin states can be stabilized below the
cooperative paramagnetic regime, such as spin liquid, spin glass or
spin ice. In some cases, long range magnetic order is established,
either by structural distortions that lift the ground state
degeneracy, or by the "order-by-disorder" mechanism\cite{Villain}.
\Kag\ antiferromagnets have been widely studied, in particular
because a spin-liquid state is predicted at low temperatures for the
S=1/2 Heisenberg system. However, real \Kag\ lattices display either
spin glass behavior or long range order with the well known
propagation vectors k=0 and k=($\frac{1}{3},\frac{1}{3}$) (often
referred to as k=($\sqrt{3}$,$\sqrt{3}$))
structures\cite{Jarosite1,Jarosite2}. More exotic ground states,
including incommensurate spin density waves and cycloidal
structures, have been found in related lattices
such as the so-called Kagome-staircase\cite{KagomeStaircase}.\\
\indent Recently, a new class of geometrically frustrated magnets,
with formula \REBa\ (R=Y,rare-earth ion) has been reported
\cite{Valldor1,Valldor2,Valldor3}. The crystal structure is built up
of alternating \Kag\ and triangular cobalt lattices, in a similar
way as SrCr$_{9x}$Ga$_{12-9x}$O$_{19}$ (SCGO)\cite{SCGO}, but with
magnetic Co$^{2+}$/Co$^{3+}$ ions in a 3:1 ratio on both
crystallographic sites (S=1.625 on average). The magnetic network in
\REBa\ is reminiscent of that of hexagonal ice 1h \cite{Ice1H}, but
with one important difference: here, half of the triangles in the
\Kag\ sheets are \emph{bi-capped} by Co atoms of the adjacent
triangular layers, providing magnetic super-exchange interactions
along the third direction.  An added element of interest is that
this network is built of corner sharing CoO$_{4}$ tetrahedra, as
opposed to the more common octahedral or pyramidal arrangement of
other frustrated magnetic oxides. The ground state of this new
topology is yet unknown, although one would suspect that the
inter-planar coupling in the third direction might play an important
role in lifting the geometrical frustration. Recently, Valldor
\cite{Valldor3} reported magnetic susceptibility measurements of
\YBa\ that indicate strong AFM interactions, with an extracted
Curie-Weiss temperature $\Theta$= -907K, and a spin-glass transition
T$_f$=66K, yielding a "frustration index" $\Theta$/T$_f$ $\sim$ 14.
More recent single-crystal neutron diffraction measurements by Soda
et al.\cite{Soda} have identified the presence of diffuse magnetic
scattering at low temperatures, suggesting short range magnetic
correlations. Broad magnetic scattering below 105K is observed at
two inequivalent positions in reciprocal space, which is interpreted
as arising from \emph{independent} short-range ordering
of both triangular and \Kag\ lattices. \\
\indent In a previous paper on isostructural \YbBa, we have
established a clear link between oxygen content and magnetic
properties, showing that only N$_2$ annealed samples display
long-range magnetic order\cite{Ashfia1}.  However, the details of
the spin ordering pattern are difficult to establish for the Yb
compound, due to the presence of two distinct propagation vectors
and strong microstructural effects. In this Letter, we report that
an\emph{oxygen stoichiometric} sample of \YBa\ undergoes a
transition to a long range ordered magnetic state below 105K, with a
single propagation vector \textbf{k}=0. Using neutron powder
diffraction data, we have determined the complex spin arrangement
that results from the competition between in-plane interactions in
the \Kag\ lattice, and AFM interactions along the $c$-axis. The
magnetic structure is strongly temperature dependent, with a gradual
spin reorientation above 30K, indicating the comparable strengths of
the in-plane and out-of-plane interactions.  Short-range magnetic
correlations are present in an extended temperature range above
110K, associated with a magnetic susceptibility behavior typical of
frustrated systems.
\\
\indent \YBa\ was synthesized from high purity Y$_2$O$_3$, BaCO$_3$,
and Co$_3$O$_4$. Repeated firing in air at 1150\Deg\ C yielded an
oxygen hyperstoichiometric compound, YBaCo$_4$O$_{7+x}$, with
x$\sim$0.2, as determined by thermogravimetry. X-ray diffraction
shows this to be biphasic, containing oxygen-rich and oxygen poor
phases. Treating this as-made material in nitrogen at 900\Deg\ C
removed the excess oxygen to produce a single-phase material with
x=-0.05(5), as we recently showed for the Yb analog\cite{Ashfia1}.
Magnetic susceptibility was collected under a magnetic field of 1
Tesla, on warming between 2K and 300K at a sweep rate of 2K/min,
using a Vibrating Sample Magnetometer (Quantum Design, PPMS).
Medium-resolution neutron powder diffraction data were collected on
warming using the GEM diffractometer of the ISIS facility
(Rutherford Appleton Laboratory) equipped with a He cryostat.
Additional high resolution diffraction patterns were recorded on the
HRPD diffractometer and the long-wavelength OSIRIS diffractometer.
Rietveld analysis was carried out using the FullProF
software\cite{Fullprof}. High resolution neutron patterns were used
to determine the ordered magnetic structures. A Simulated Annealing
procedure\cite{Fullprof} including 134 integrated intensities,
regrouped in 36 clusters of overlapping reflections, was employed
to explore the possible spin arrangements\\
\begin{figure}[h!]
\includegraphics[scale=0.7]{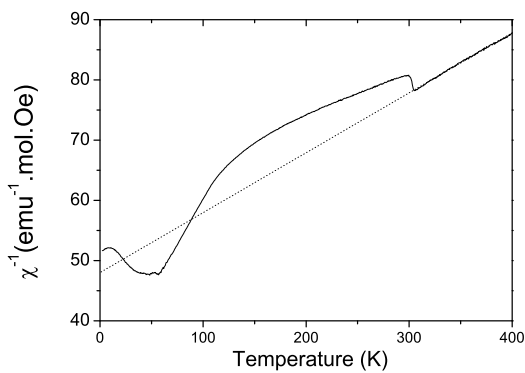}
\caption{Inverse magnetic susceptibility of \YBa\ collected at 1000
Oe. The dotted line is a fit of the high temperature data with a
Curie-Weiss law.}
\end{figure}
\indent The reciprocal magnetic susceptibility, reported in Fig. 1,
shows a Curie-Weiss behavior above 313K, with a Curie constant
C=10.08 emu.mol$^{-1}$K$^{-1}$ and a Weiss temperature
$\Theta_{CW}$=-508K indicating dominant antiferromagnetic (AFM)
interactions.  In the linear high-temperature region the system is
best described by a cooperative paramagnetic regime down to
T$<<$$\Theta_{CW}$, a signature of geometrically frustrated
systems\cite{Schiffer}.  The effective moment per formula unit is
8.9 $\mu_B$, in good agreement with the calculated spin-only value
of 8.3 $\mu_B$ considering a mixture of 3Co$^{2+}$/1Co$^{3+}$. Below
313K an abrupt jump in the magnetic susceptibility is observed (Fig
1) as a result of a first-order structural phase transition, similar
to the Yb analog \cite{Ashfia1}.  High resolution neutron
diffraction data confirm that \YBa\ has trigonal symmetry (space
group $P3_1c$) at high temperature and undergoes a structural phase
transition to orthorhombic symmetry, space group $Pbn2_1$ with an
(a,$\sqrt{3}$a,c) supercell, below T$_s$=313K. The transition occurs
as a response to a strongly underbonded Ba ion and does not appear
to involve charge ordering on the Co ions\cite{Ashfia1}. At low
temperatures, \YBa\ displays a slight monoclinic, distortion, as
clearly evidenced by our high-resolution neutron data (see below) as
well as by preliminary synchrotron data [17].  A detailed study of
these phase transitions, outside the scope of this letter, will be
published elsewhere. We note however that the structural transition
directly influences the position of oxygens in the crystal and
induces a buckling of the CoO$_4$ tetrahedra in the \Kag\ sheets
\cite{Ashfia1}. \indent Since the local environment of the Co ions
is only slightly affected, the change of slope in $\chi^{-1}$ below
T$_s$ is probably not related to an enhancement of the effective
moment. Instead, the pronounced curvature below T$_s$ is likely to
arise from the onset of short-range correlations, as frustration is
lifted by the lowering of symmetry.\\
 \indent A signature of the growing
short-range correlations is directly evidenced in the neutron
diffraction data. Specifically, magnetic diffuse scattering appears
below 250K (inset of Fig. 2), tracking the pronounced deviation seen
in $\chi^{-1}$. This scattering has no counterpart in X-ray
synchrotron data\footnote{A. Huq, Personal communication},
indicating its purely magnetic origin, and is centered at a
wavevector $\mid$Q$\mid$=1.35\AA$^{-1}$, i.e., the same position of
the most intense antiferromagnetic Bragg peak at lower temperatures.
The exact nature of this broad magnetic feature is uncertain, due to
the weakness and breadth of the signal. In particular, it is unclear
whether it is 2D Warren scattering\cite{Warren} or arises from 3D
short range correlations. Assuming a symmetric profile of the
diffuse scattering, modeled by a Lorentzian function, we estimate a
correlation length of the short-range order $\xi$ between 10\AA\ at
200K and 35\AA\ at 115K.\\
\indent Below 115K, the diffuse scattering abruptly sharpens, as
illustrated on figure 2, giving rise to resolution-limited magnetic
reflections that can all be indexed at k=0 with respect to the
primitive orthorhombic cell (or k=($\frac{1}{2}$,0,0) in the
hexagonal setting). The appearance of resolution-limited magnetic
peaks at a single position in reciprocal space contrasts with recent
neutron diffraction experiments on single crystals of
YBaCo$_4$O$_{7.2}$ \cite{Soda}, where a broad magnetic response was
reported in the entire temperature range. Most likely the
discrepancy is attributable to the fact that the single crystal
studied by Soda \textit{et al.} is not stoichiometric in oxygen.
Indeed, our neutron powder data on an oxygen-rich, biphasic sample
of YBaCo$_4$O$_{7.22(2)}$ similarly shows broad magnetic scattering
at the same multiple Q-points.
\begin{figure}[h!]
\includegraphics[scale=0.30]{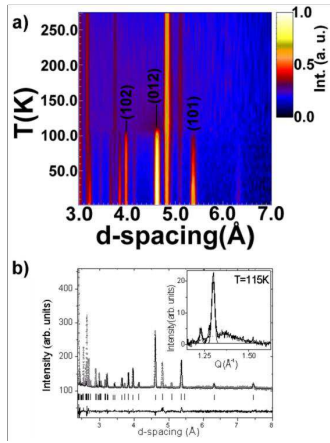}
\caption{a) Thermodiffractogram of \YBa\ collected on GEM. Data from
a bank of detectors situated at 34.96\Deg\-2$\Theta$ is shown. The
scattering intensity is color-coded and normalized to the most
intense magnetic reflection (102). b) Rietveld refinement at 2K. The
grey crosses and dashed solid line represent the experimental data
points and calculated diffraction pattern, respectively. The
difference is shown at the bottom as a solid line. The raw of
markers shows the position of the nuclear and magnetic reflection.
The thick solid line represents the contribution from magnetic
scattering alone. The inset of b) display a section of the neutron
diffraction pattern collected at 115K on GEM showing the coexistence
of nuclear Bragg scattering and magnetic diffuse scattering. The
solid line is a fit to the data. The diffuse magnetic scattering is
modeled by a Lorentzian profile function. }
\end{figure}
\begin{figure}[h!]
\includegraphics[scale=0.2]{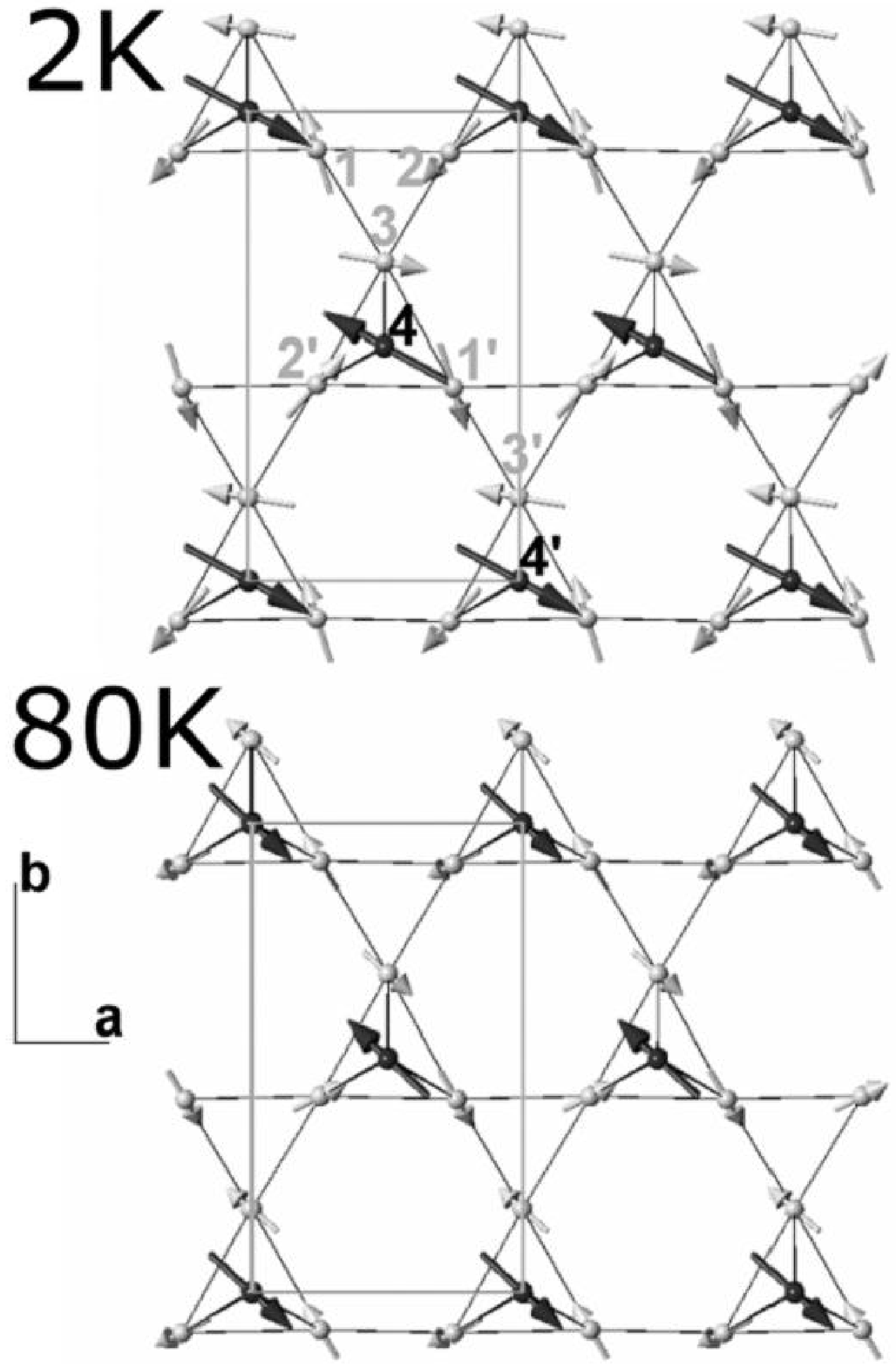}
\includegraphics[scale=0.2]{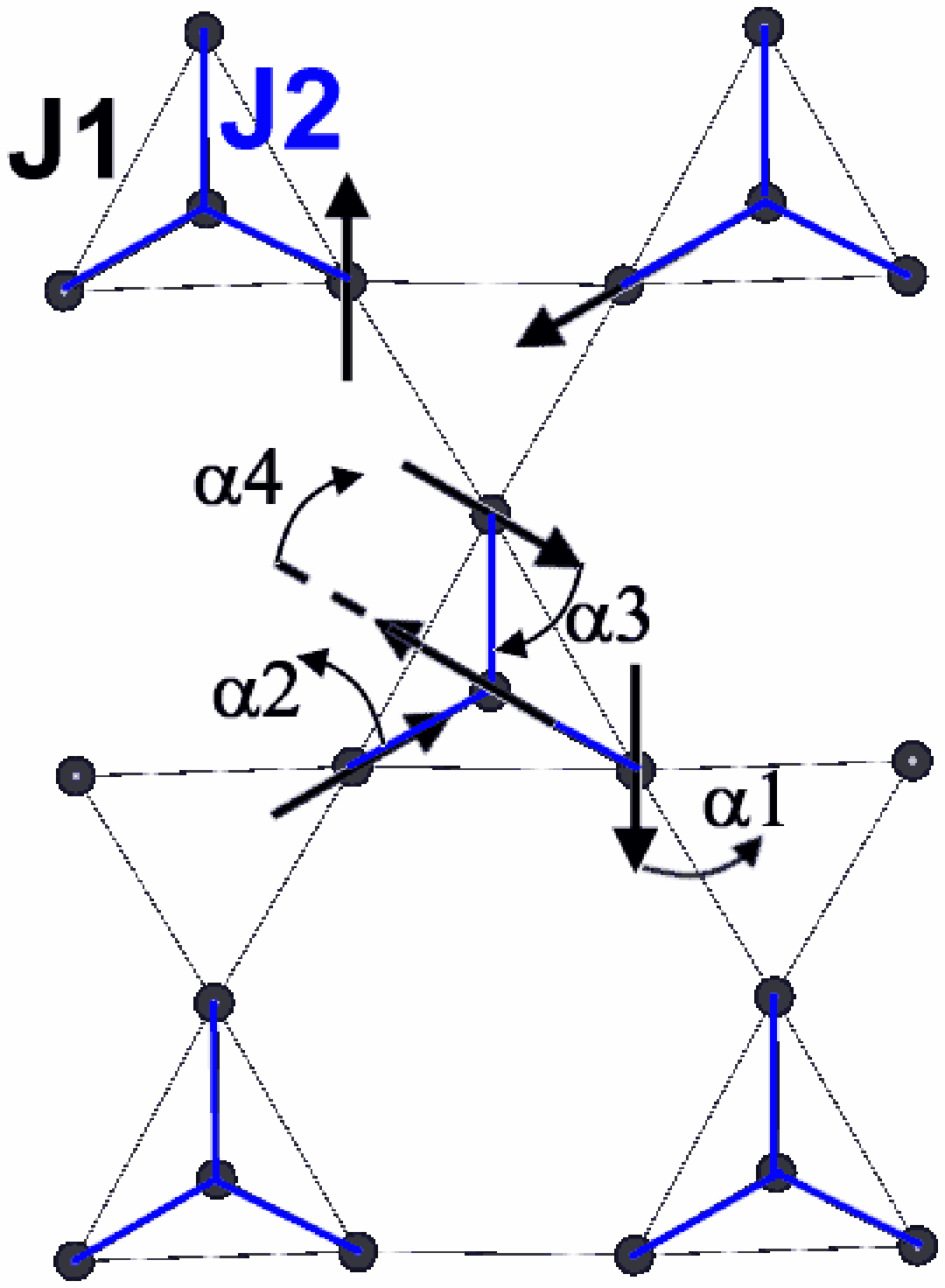}
\caption{Left panel: Magnetic structure of \YBa\ projected in the
\textit{ab}-plane at 2K (top panel) and 80K (bottom). The black and
grey arrows represent moments on the Co(1) and Co(2) sites,
respectively (see text for details). For clarity only one of the two
Co(1)/Co(2) layers is shown, i.e atoms with fractional coordinates z
$>$0.5. Magnetic moments on the second layer, not shown, are related
by time reversal of the 2$_1$ symmetry operator.Right panel:
Exchange interactions (J$_1$, J$_2$, color online) and ideal 120
\Deg\ structure (see text for details).}
\end{figure}
The magnetic structure at 2K, shown in Fig. 3, corresponds to a
simultaneous ordering of Co spins in both the \Kag\ planes and
triangular layers.  Using high-resolution neutron data, we could
determine the absolute orientation of the spins with respect to the
crystal axes, a crucial element to determine the magnetic symmetry.
The magnetic structure can be described in the Shubnikov group
\textit{P112$_1$'}, indicating a loss of both glide planes of the
orthorhombic paramagnetic group. Indeed, evidence for a monoclinic
distortion ($\gamma$=90.135(2)\Deg) is unambiguously seen in high
resolution data at high-Q. Because of this lower symmetry, all the
spin configurations are admixtures of two irreducible
representations of Pbn2$_1$.  The results of Simulated Annealing
runs invariably lead to a supplementary set of constraints that are
not imposed by symmetry: (1) all spins lie approximatively in the
\textit{ab}-plane; (2) spins labeled 1-4 in Fig. 3 are found
antiparallel to their primed counterparts (1'-4'). Rietveld
refinements with this model fit
the data well throughout the entire temperature range from 2K to 100K.\\
\indent At 2K, the refined value of the magnetic moment on the
triangular layer is 3.49(8)$\mu_B$, in close agreement with the
expected 3.25$\mu_B$ assuming spin-only contribution from a Co ion
with intermediate valence 2.25. In the \Kag\ planes, however, the
moment (2.19(4)$\mu_B$) is substantially reduced, revealing that a
large fraction of the spins remain disordered at low temperatures.
The three-dimensional magnetic structure can be described as
follows: The spins on non-capped triangular units of the \Kag\
sheets adopt configuration with total moment \textbf{M$_1$}$\simeq$
0, corresponding to one of the negative-chirality arrangement found
in the well-known k=0 and k=($\frac{1}{3}$,$\frac{1}{3}$)
structures. Spins belonging to bi-capped triangular units have a
total moment \textbf{M$_2$}$\neq$0, \textit{antiparallel} to the
Co(4) spins capping the triangle above and below, resulting in
antiferromagnetic chains running along \textit{c}. The magnitudes of
\textbf{M$_2$} and the Co(4) moment are different, implying that
each individual column is a ferrimagnetic entity. Globally, the 3D
structure does not carry a net moment, as
adjacent chains are oriented antiparallel.\\
\indent The ordered moment extracted from Rietveld refinements,
shown in Fig.4 a) varies smoothly with temperature, but the
temperature evolution of selected magnetic Bragg peaks (Fig. 4b)
indicates a gradual reorientation between 30 and 60K. For example,
the (012) peak intensity departs from mean-field behavior at 60K,
correlating with the minimum in $\chi^{-1}$. We have parameterized
the spin reorientation by 4 angles $\alpha_i$ (i=1-4), which measure
the deviation of individual spins from the ideal 120\Deg\ structure
shown in Fig. 3b. The low temperature magnetic arrangement is within
a few degrees of the ideal model. On warming above 30K, the
deviation becomes more pronounced with a tendency towards saturation
above 60K. In the high temperature structure (T$>$60K), spins on
Co(1), Co(3)and Co(4) are essentially \emph{collinear} whereas the
spin on Co(2) becomes approximately \emph{orthogonal} to the
others.\\
\begin{figure}[h!]
\includegraphics[scale=0.6]{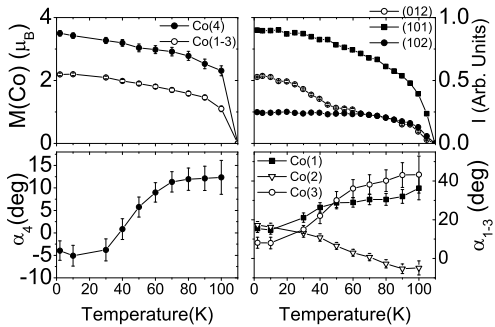}
\caption{a) Temperature dependence of the magnetic moment on the
Co(1) and Co(2-4) sites. b) Integrated intensities of selected
magnetic peaks (labels ) versus temperature. c-d) Angular deviations
from the ideal 120\Deg\ model for Co ions labeled 1-4 (see Fig. 3
and text for details.)}
\end{figure}
\indent Within the framework of a simple nearest-neighbor isotropic
exchange model, the evolution from the 120\Deg\ to the collinear
structures can be understood in terms of competition between the
in-plane interaction (J$_1$ in Fig. 3c), which favors the 120\Deg\
structure and the out-of-plane interaction J$_2$, which tends to
stabilize collinear AFM chains along the c-axis.
 However, the orthogonal orientation of the Co(2) spin is difficult to reproduce with such a simple model.
A straightforward parametrization of the observed high- and
low-temperature structure, the former including an angular parameter
$\theta$ to describe the orientation of the Co(2) spin, leads to
energy expressions:
\begin{equation}
E_{HT}(\theta)=-2J_2S_1S_2+(S_1)^2sin(\theta)[\frac{S_2}{S_1}.J_2-2J_1]
\end{equation}
\begin{equation}
E_{LT}=-J_1S_1^2-2J_2S_1S_2
\end{equation}
At high temperature, the minimum energy configuration is obtained
for either a collinear or an anti-collinear structure, depending on
the ratio J$_1$S$_1$/J$_2$S$_2$.  The observed 90 \Deg arrangement
of the Co(2) spin must therefore involve additional exchange or
single ion anisotropy terms. By varying the ratio S$_1$/S$_2$ for a
given set of exchange parameters, one finds a crossover between the
collinear (or anticollinear) and LT structures.   \\
\indent In summary, a first-order structural phase transition at
313K relieves the geometric frustration in \YBa\, giving rise to an
extended regime of short range correlations followed by 3D
antiferromagnetic ordering below 110K. All Co-Co interactions are
found to be antiferromagnetic, in agreement with the large negative
Weiss temperature. The low-temperature magnetic structure is related
to the 120\Deg\ structures commonly found in \Kag\ systems, with S=0
arrangement of spins in uncapped triangular units of the \Kag\
sheets.  Capped triangular units carry a net spin oriented
antiparrallel to that of the capping Co ions.  On warming, spin
reorientation leads to progressive loss of the the S=0, 120\Deg\
arrangement, favoring the formation of ferrimagnetic collinear
chains along the c-axis and suggesting similar energy scales for the
competing in-plane and out-of-plane interactions. As the lifting of
geometrical frustration is very sensitive to the structural
distortion, we suggest that the magnetic phase diagram of \REBa\
could be tuned by chemical substitutions or the application of
external perturbations.

%\bibliography{RBaCo4O7}

\end{document}